\documentclass[twovoluments,showpacs,preprintnumbers]{revtex4}
\usepackage{amssymb}
\usepackage{amsmath}
\usepackage{txfonts}
\usepackage[mathscr]{eucal}

\input{tcilatex}
\begin{document}

\title{Landauer transport model for Hawking radiation from a
Reissner-Nordstrom black hole}
\author{Xiao-Xiong Zeng}
\affiliation{Department of Physics, Institute of Theoretical Physics, Beijing Normal
University, Beijing, 100875, China}
\author{Shi-Wei Zhou}
\affiliation{Department of Physics, Institute of Theoretical Physics, Beijing Normal
University, Beijing, 100875, China}
\author{Wen-Biao Liu (corresponding author)}
\email{wbliu@bnu.edu.cn}
\affiliation{Department of Physics, Institute of Theoretical Physics, Beijing Normal
University, Beijing, 100875, China}

\begin{abstract}
The recent work of Nation et al in which Hawking radiation energy and
entropy flow from a black hole can be regarded as a one-dimensional (1D)
Landauer transport process is extended to the case of a Reissner-Nordstrom
(RN) black hole. It is found that the flow of charge current can also be
transported via a 1D quantum channel except the current of Hawking
radiation. The maximum entropy current, which is shown to be particle
statistics independence, is also obtained.

\end{abstract}

\maketitle

\section{Introduction}

Hawking radiation from a black hole is an eternal topic in theoretical
physics because it provides not only a clue to detect black holes but also a
platform to research the quantum gravity. Since the first proposal of
Hawking \cite{1,2} that a black hole can emit radiation in the curved space
time background, there are many methods to derive it \cite{3,4,5,6,7,8,9}.
Now it is believed that Hawking radiation arises from the production of
virtual particle pairs spontaneously near the inner of horizon due to the
vacuum fluctuation. When the negative energy virtual particle tunnels
inwards, the positive energy virtual particle materializes as a real
particle and escapes to infinity \cite{10,11,12,13,14,15,16,17,18,19,20,21}.
But, how does the positive energy particle run? Recently, Nation et al \cite%
{22} gave a model where the positive energy particle escapes to infinity via
a 1D quantum channel. The key idea there was that the black hole and vacuum
can be viewed as two thermal reservoirs connected by a single 1D quantum
channel. As the scattering effect was ignored, Hawking radiation flow rate
was obtained using Landauer transport model, which is shown to be equal to
the energy-momentum tensor expectation value of an infinite observer as the
left and right are regarded as the black hole and the thermal environment
with absolute temperature zero surrounding the black hole. They also found
that the upper limits of entropy current were same for both bosons or
fermions.

In this letter, we would like to extend this idea to study Hawking radiation
of charged particles from a RN black hole. When the near horizon conformal
symmetry is considered, the expectation value of electric current flow must
be investigated except the flow of energy-momentum tensor due to the
existence of electromagnetic field. In Landauer transport model, the flow of
charge current must also be considered besides the energy and entropy flows.
In the letter, we use the units $G=\hbar =c=k_{B}=1$.

\section{Conformal symmetry and Hawking radiation flux}

The line element of a RN black hole is%
\begin{equation}
ds^{2}=-fdt^{2}+f^{-1}dr^{2}+r^{2}d\theta ^{2}+r^{2}sin^{2}\theta d\varphi
^{2},  \label{1}
\end{equation}%
where%
\begin{equation}
f=1-\frac{2M}{r}+\frac{Q^{2}}{r},  \label{2}
\end{equation}%
in which $M,Q$ are the mass and charge of the black hole. The
electromagnetic four-vector is%
\begin{equation}
A_{\mu }=(-\frac{Q}{r},0,0,0).  \label{3}
\end{equation}%
From Eq.(\ref{1}), one can get the event horizon $r_{h}=M+\sqrt{M^{2}-Q^{2}}$
immediately.

According to the dimensional reduction technique, Eq.(\ref{1}) can be
expressed as the two-dimensional form effectively%
\begin{equation}
ds^{2}=-f(r)dt^{2}+f(r)^{-1}dr^{2}.  \label{4}
\end{equation}%
Under the tortoise coordinate transformation defined as $dr^{\ast }=\frac{1}{%
f(r)}dr$ and the null coordinates $u=t+r^{\ast }$, $v=t-r^{\ast },$ we can
construct the Kruskal coordinates as $U=-\frac{1}{\kappa }e^{-{\kappa u}}$, $%
V=\frac{1}{\kappa }e^{\kappa v}$, so we get the corresponding conformal form%
\begin{equation}
ds^{2}=-f(r)dudv,  \label{5}
\end{equation}%
\begin{equation}
ds^{2}=-f(r)e^{-2{\kappa r^{\ast }}}dUdV,  \label{6}
\end{equation}%
in which $\kappa =\frac{r_{h}-M}{r_{h}^{2}}$ is the surface gravity.

In the following, we intend to find the expectation values of
energy-momentum tensor and gauge current. It is well known that the
classical Einstein field equation can be derived from the classical action
by the minimal variational principle. In the semiclassical theory, this
principle is still valid and thus we have%
\begin{equation}
\frac{2}{\sqrt{-g}}\frac{\delta \Gamma }{\delta g^{\mu \nu }}=\langle T_{\mu
\nu }\rangle ,  \label{7}
\end{equation}%
in which $\Gamma $ is the effective action with central charge $c=1$. In the
gravitational field with electric field background, the action consists the
contributions of gravitational part \cite{23,24,25,26,27,28}%
\begin{equation}
\Gamma _{grav}=\frac{1}{96\pi }{\int d^{2}xd^{2}y\sqrt{-g}R(x)\frac{1}{%
\Delta _{g}}(x,y)\sqrt{-g}R(y)},  \label{8}
\end{equation}%
and the gauge part%
\begin{equation}
\Gamma _{U(1)}=\frac{e^{2}}{2\pi }{\int d^{2}xd^{2}y\epsilon ^{\mu \nu
}\partial _{\mu }A_{\nu }(x)\frac{1}{\Delta _{g}}(x,y)\epsilon ^{\rho \sigma
}\partial _{\rho }A_{\sigma }(y)},  \label{9}
\end{equation}%
in which $R$ is the two-dimensional scalar curvature, $\Delta _{g}$ is the
Laplacian, $\epsilon ^{\mu \nu }$ represents the two-dimensional Levi-Civita
tensor. From these actions, the expectation values of energy-momentum tensor
and gauge current can be solved as \cite{25,26,27,28}%
\begin{equation}
\langle T_{\mu \nu }^{grav}\rangle =\frac{1}{48\pi }({2g_{\mu \nu }R-2\nabla
_{\mu }\nabla _{\nu }S+\nabla _{\mu }S\nabla _{\nu }S-\frac{1}{2}g_{\mu \nu
}\nabla ^{\rho }S\nabla _{\rho }S}),  \label{10}
\end{equation}%
\begin{equation}
\langle T_{\mu \nu }^{U(1)}\rangle =\frac{e^{2}}{\pi }({\nabla _{\mu
}B\nabla }_{\nu }B{-\frac{1}{2}g_{\mu \nu }\nabla ^{\rho }B\nabla _{\rho }B}%
),  \label{11}
\end{equation}%
\begin{equation}
\langle J^{\mu }\rangle =\frac{1}{\sqrt{-g}}\frac{\delta \Gamma }{\delta
A_{\mu }}=\frac{e^{2}}{\pi }\frac{1}{\sqrt{-g}}\epsilon ^{\mu \nu }\partial
_{\nu }B,  \label{12}
\end{equation}%
where%
\begin{equation}
S(x)=\int {d^{2}y\frac{1}{\Delta _{g}}(x,y)\sqrt{-g}R(y)},  \label{13}
\end{equation}%
\begin{equation}
B(x)=\int {d^{2}y\frac{1}{\Delta _{g}}(x,y)\epsilon ^{\mu \nu }\partial
_{\mu }A_{\nu }(y)}.  \label{14}
\end{equation}%
The concrete values of above equations have been given in Refs.\cite{25,26},
where the boundary conditions are invoked. In order to get Hawking radiation
flux, we must use boundary conditions as Unruh effect in the Unruh vacuum.
In the advanced Eddington coordinate, there is a finite mount of flux at the
horizon while no flux at infinity. Meanwhile, in the retarded Eddington
coordinate, there is no flux at the horizon while a finite mount of flux at
infinity. For observers in the Kruskal coordinate, the relation $J_{U}=-%
\frac{J_{u}}{\kappa U}$ and $T_{UU}=(\frac{1}{\kappa U})^{2}T_{uu}$ also
should be imposed \cite{23,24}. In this case, we find the gauge fluxes%
\begin{equation}
\langle J_{u}\rangle =\frac{e^{2}}{2\pi }(A_{t}(r)-A_{t}(r_{h})),  \label{15}
\end{equation}%
\begin{equation}
\langle J_{v}\rangle =\frac{e^{2}}{2\pi }A_{t}(r),  \label{16}
\end{equation}%
and the energy-momentum tensor fluxes%
\begin{equation}
\langle T_{uu}\rangle =\frac{1}{24\pi }(\frac{{\kappa _{h}^{2}}}{2}-\frac{M}{%
r^{3}}+\frac{3M^{2}}{2r^{4}}+\frac{3e^{2}}{2r^{4}}-\frac{3Me^{2}}{r^{5}}+%
\frac{e^{4}}{r^{6}})+\frac{e^{2}}{4\pi }(A_{t}(r)-A_{t}(r_{h}))^{2},
\label{17}
\end{equation}%
\begin{equation}
\langle T_{vv}\rangle =\frac{1}{24\pi }(-\frac{M}{r^{3}}+\frac{3M^{2}}{2r^{4}%
}+\frac{3e^{2}}{2r^{4}}-\frac{3Me^{2}}{r^{5}}+\frac{e^{4}}{r^{6}})+\frac{%
e^{2}}{4\pi }A_{t}^{2}(r).  \label{18}
\end{equation}%
Obviously, from Eq.(\ref{17}) we can get that the infinite observers will
see a bunch of flow $\frac{\pi {T_{h}^{2}}}{12}+\frac{e^{2}}{4\pi }%
A_{t}^{2}(r_{h})$, in which $T_{h}=\frac{\kappa }{2\pi }$ is Hawking
temperature. Correspondingly, Eq.(\ref{18}) indicates that a near horizon
observer may find that a bunch of flow $-\frac{\pi {T_{h}^{2}}}{12}+\frac{%
e^{2}}{4\pi }A_{t}^{2}(r_{h})$ will drop into event horizon. From Eq.(\ref%
{15}) and Eq.(\ref{16}), we also know that a negative flow $\frac{e^{2}}{%
2\pi }A_{t}(r_{h})$ at the event horizon is responsible for the flow of
charge current observed by an infinite observer.

\section{Hawking radiation from Landauer transport model}

The Landauer transport model was first proposed to study electrical
transport in mesoscopic circuits, and subsequently used to describe thermal
transport. In 2000, the phononic quantized thermal conductance counterpart
was measured for the first time \cite{29}. For the 1D quantum channel of
thermal conductance, it is supposed that there are two thermal reservoirs
characterized by the temperatures $T_{L}$, $T_{R}$ and chemical potentials $%
u_{L}$, $u_{R}$, where subscripts L and R denote the left and right thermal
reservoirs respectively. They are coupled adiabatically through an effective
1D connection. Because of the temperature difference, the thermal
transportation will happen. Typically, a wire will provide several available
parallel channels for the given values of chemical potential and
temperature. For the sake of simplicity in the present investigation, we
restrict only to ballistic transport, which means the channel currents do
not interfere with each other \cite{22}. Universally, there are several
distribution functions and here we adopt the Haldane's statistics \cite{30}%
\begin{equation}
f_{g}(E)=\{ \omega \lbrack \frac{(E-u)}{T}]+g\}^{-1},  \label{19}
\end{equation}%
where the function $\omega (x)$ satisfies the relation%
\begin{equation}
\omega (x)^{g}[1+\omega (x)]^{1-g}=e^{x},  \label{20}
\end{equation}%
in which $g=0$, $g=1$ correspond bosons and fermions respectively.

The left (right) components of the single channel energy and entropy
currents are%
\begin{equation}
\dot{E}_{L(R)}=\frac{T_{L(R)}^{2}}{2\pi }\int_{x_{L(R)}^{0}}^{\infty }dx(x+%
\frac{u_{L(R)}}{T_{L(R)}})f_{g}(x),  \label{21}
\end{equation}%
\begin{eqnarray}
\dot{S}_{L(R)} &=&\frac{T_{L(R)}}{2\pi }\int_{x_{L(R)}^{0}}^{\infty
}dx\{f_{g}ln{f_{g}}+(1-gf_{g})ln{(1-gf_{g})}  \notag \\
&&-[1+(1-gf_{g})]ln[{1+(1-g)f_{g})}]\},  \label{22}
\end{eqnarray}%
in which $x_{L(R)}^{0}=-\frac{u_{L(R)}}{T_{L(R)}}$. As done in Ref.\cite{22}%
, we define the zero level of energy with respect to the longitudinal
component of the kinetic energy. The total energy and entropy currents are
then just $\dot{E}=\dot{E}_{L}-\dot{E}_{R}$, $\dot{S}=\dot{S}_{L}-\dot{S}%
_{R} $. We first consider the case of fermions, namely $g=1$. According to
the viewpoint of Landauer, the maximum energy flow rate of fermions can be
treated as the combination of fermionic particle and antiparticle single
channel currents. As $x=\frac{(E-u)}{T}$, $y=\frac{(E+u)}{T}$ are defined,
the flow of energy therefore can be expressed as%
\begin{equation}
\dot{E}_{L(R)}=\frac{T_{L(R)}^{2}}{2\pi }\bigg[\int_{\frac{-u_{L(R)}}{%
T_{L(R)}}}^{\infty }dx(x+\frac{u_{L(R)}}{T_{L(R)}})\frac{1}{e^{x}+1}+\int_{%
\frac{u_{L(R)}}{T_{L(R)}}}^{\infty }dy(y+\frac{u_{L(R)}}{T_{L(R)}})\frac{1}{%
e^{y}+1}\bigg].  \label{23}
\end{equation}

To get the final value, we first vary the lower limit of integral, that is%
\begin{eqnarray}
\dot{E}_{L(R)} &=&\frac{T_{L(R)}^{2}}{2\pi }\bigg[\int_{0}^{\infty }dx(x+%
\frac{u_{L(R)}}{T_{L(R)}})\frac{1}{e^{x}+1}+\int_{0}^{\frac{u_{L(R)}}{%
T_{L(R)}}}dx(-x+\frac{u_{L(R)}}{T_{L(R)}})\frac{1}{e^{-x}+1}+  \notag \\
&&\int_{0}^{\infty }dy(y+\frac{u_{L(R)}}{T_{L(R)}})\frac{1}{e^{y}+1}%
-\int_{0}^{\frac{u_{L(R)}}{T_{L(R)}}}dy(y+\frac{u_{L(R)}}{T_{L(R)}})\frac{1}{%
e^{y}+1}\bigg].  \label{24}
\end{eqnarray}%
As $\frac{1}{e^{-x}+1}=1-\frac{1}{e^{x}+1}$ is replaced, Eq.(\ref{24}) takes
the form as%
\begin{eqnarray}
\dot{E}_{L(R)} &=&\frac{T_{L(R)}^{2}}{2\pi }\bigg[\int_{0}^{\infty }\frac{x}{%
e^{x}+1}dx+\int_{0}^{\infty }\frac{y}{e^{y}+1}dy+2\frac{u_{L(R)}}{T_{L(R)}}%
(\int_{0}^{\infty }\frac{1}{e^{x}+1}dx  \notag \\
&&-\int_{0}^{\frac{u_{L(R)}}{T_{L(R)}}}\frac{1}{e^{x}+1}dx)+\frac{%
u_{L(R)}^{2}}{2T_{L(R)}^{2}}\bigg].  \label{25}
\end{eqnarray}

According to the technique of Landau \cite{31}, the convergence rate of $%
\frac{1}{e^{x}+1}$ is very fast, the upper limit of integral $\frac{u_{L(R)}%
}{T_{L(R)}}$ hence can be changed as infinity. So we get the final value of
the energy current%
\begin{equation}
\dot{E}=\dot{E}_{L}-\dot{E}_{R}=\frac{\pi }{12}({T_{L}^{2}-T_{R}^{2}})+\frac{%
1}{4\pi }({u_{L}^{2}-u_{R}^{2}}).  \label{26}
\end{equation}%
As the left and right are regarded as the RN black hole and the thermal
environment with absolute temperature zero surrounding the black hole
respectively, we find%
\begin{equation}
\dot{E}=\frac{\pi {T_{h}^{2}}}{12}+\frac{e^{2}}{4\pi }A_{t}^{2}(r_{h}),
\label{27}
\end{equation}%
in which $u_{R}=0$ and $u_{L}=\frac{eQ}{r_{h}}$ are used. Obviously, this
result agrees with the energy-momentum tensor flux obtained by conformal
symmetry.

For the entropy flow in Eq.(\ref{22}), the value is independent on the
chemical potential and it depends only on the lower limit of integral. We
can get its maximum value after calculating, $\dot{S}=\frac{\pi {T_{h}}}{6}$%
, under the degenerate limit $\frac{u_{L}}{T_{L}}\rightarrow \infty $. As
the maximum energy and entropy current expressions are combined by
eliminating $T_{L}$, the relation $\dot{S}^{2}\leq \frac{\pi {\dot{E}}}{3}$,
which was proved to be universal for 1D quantum channels with arbitrary
reservoir temperatures, chemical potentials and particle statistics \cite%
{22,32,33}, also can be reproduced.

Thinking of the transportation of charge via the 1D quantum channel, the
current flow from left (with higher chemical potential $u$) to right without
scattering can be expressed as \cite{32}%
\begin{equation}
\dot{I}=\frac{e}{2\pi }\int_{0}^{\infty }f(\omega )d\omega .  \label{28}
\end{equation}%
For the case of fermions, the contribution of antiparticle also should be
considered, we have%
\begin{equation}
\dot{I}=\frac{e}{2\pi }\int_{0}^{\infty }(\frac{1}{e^{\frac{\omega -u_{h}}{%
T_{h}}}+1}-\frac{1}{e^{\frac{\omega +u_{h}}{T_{h}}}+1})d\omega ,  \label{29}
\end{equation}%
in which $f(\omega )=1/({e^{\frac{\omega -u_{h}}{T_{h}}}+1})$ is the
radiation spectrum of fermions at the event horizon. Finishing the integral,
we obtain%
\begin{equation}
\dot{I}=-\frac{e^{2}}{2\pi }A_{t}(r_{h}),  \label{30}
\end{equation}%
which is consistent with the value of charge current flow at the event
horizon in Eq.(\ref{15}). Namely the gauge flux with respect to electric
charge of the RN space time also can be transported via the 1D quantum
channel.

Now, we check whether the Landauer transport model is valid for the bosons.
Putting $g=0$ into Eq.(\ref{19}) and Eq.(\ref{20}), the energy current can
be expressed as%
\begin{equation}
\dot{E}_{L(R)}=\frac{T_{L(R)}^{2}}{2\pi }\int_{\frac{-u_{L(R)}}{T_{L(R)}}%
}^{\infty }dx(x+\frac{u_{L(R)}}{T_{L(R)}})\frac{1}{e^{x}-1}.  \label{31}
\end{equation}%
After varying the lower limit of integral and adopting $\frac{1}{e^{-x}-1}%
=-1-\frac{1}{e^{x}-1}$, the above equation can be rewritten as%
\begin{equation}
\dot{E}_{L(R)}=\frac{T_{L(R)}^{2}}{2\pi }\bigg[\int_{0}^{\infty }(x+\frac{%
u_{L(R)}}{T_{L(R)}})\frac{1}{e^{x}-1}dx-\int_{0}^{\frac{u_{L(R)}}{T_{L(R)}}}(%
{\frac{u_{L(R)}}{T_{L(R)}}-x})\frac{1}{e^{x}-1}dx-\frac{u_{L(R)}^{2}}{%
2T_{L(R)}^{2}}\bigg].  \label{32}
\end{equation}%
Finishing the integration, we find%
\begin{equation}
\dot{E}_{L(R)}=\frac{T_{L(R)}^{2}}{2\pi }\bigg[\frac{{\pi }^{2}}{6}%
+\int_{0}^{\frac{u_{L(R)}}{T_{L(R)}}}ln{(e^{x}-1)}dx\bigg].  \label{33}
\end{equation}%
Adopting the similar technique as done in the case of fermions, where $\frac{%
u_{L(R)}}{T_{L(R)}}\gg 0$ can be produced, the second term of above equation
can be integrated easily. When the the left and right are treated as the RN
black hole and thermal environment with absolute temperature zero, we find%
\begin{equation}
\dot{E}=\frac{\pi {T_{h}^{2}}}{12}+\frac{e^{2}}{4\pi }A_{t}^{2}(r_{h}).
\label{34}
\end{equation}%
This result also agrees with the energy-momentum tensor flux observed by the
infinite observers in Eq.(\ref{17}). For a boson, the maximum value of
entropy flow, $\dot{S}=\frac{\pi {T_{h}}}{6}$, and the charge current flow
can also be reproduced by adopting similar skills as the case of fermions.

\section{Discussion and conclusion}

Flows of Hawking radiation energy and charge current from a RN black hole
are investigated using a 1D quantum transport model, which are shown to be
consistent with the vacuum expectation values of energy-momentum tensor and
gauge flux. The maximum value of entropy flow in different degeneration is
obtained and the result is independent on the statistics. For flows of
energy and charge current of fermions, we get the same result as that of
bosons when the contribution of antiparticle is considered, which confirms
the viewpoint of Davies \cite{33} that the fermionic field describing a
massless particle plus its antiparticle is equivalent to a single massless
bosonic field in a (1+1)-dimensional curved spacetime.

Note that when we investigate the flow of fermions and bosons using Landauer
transport model, the condition $\frac{u_{L(R)}}{T_{L(R)}}\gg 0$ should be
imposed in both of them. In statistics physics, this condition should be
satisfied in order to avoid the degeneration of boson distribution and
fermion distribution to the Boltzmann distribution.

As the first law of black hole thermodynamics and energy conservation are
considered, one also can get the net entropy production rate defined as $R=%
\frac{dS}{dS_{BH}}$ \cite{34} in the two dimensional space time. Because of
the existence of electromagnetic field, we find the production rate of RN
black hole is smaller than the case of Schwarzschild black hole with $R=2$.

\begin{acknowledgments}
This research is supported by the National Natural Science Foundation of
China (Grant Nos. 10773002, 10875012). It is also supported by the
Scientific Research Foundation of Beijing Normal University under Grant No.
105116.
\end{acknowledgments}


\begin{thebibliography}{99}
\bibitem{1} S. W. Hawking, Nature 248 (1974) 30.

\bibitem{2} S. W. Hawking, Commun. Math. Phys. 43 (1975) 199.

\bibitem{3} G. Gibbons, S. W. Hawking, Phys. Rev. D 15 (1977) 2752.

\bibitem{4} A. Strominger, C. Vafa, Phys. Lett. B 99 (1996) 379.

\bibitem{5} G. W. Gibbons, M. J. Perry, Proc. Roy. Soc. Lond. A 358 (1978)
467.

\bibitem{6} M. K. Parikh, F. Wilczek, Phys. Rev. Lett. 85 (2000) 5042.

\bibitem{7} W. G. Unruh, Phys. Rev. D 14 (1976) 870.

\bibitem{8} T. Damour, Phys. Rev. D 18 (1978) 18.

\bibitem{9} S. P. Robinson, F. Wilczek, Phys. Rev. Lett. 95 (2005) 011303.

\bibitem{10} J. Zhang, Z. Zhao, Nucl. Phys. B 725 (2005) 173.

\bibitem{11} J. Zhang, Z. Zhao, JHEP 0510 (2005) 055.

\bibitem{12} W. B. Liu, Phys. Lett. B 634 (2006) 541.

\bibitem{13} S. Z. Yang, Chin. Phys. Lett. 22 (2005) 2492.

\bibitem{14} K. Xiao, W. B. Liu, H. B. Zhang, Phys. Lett. B 647 (2007) 482.

\bibitem{15} M. Alves, Int. J. Mod. Phys. D 10 (2001) 575.

\bibitem{16} E. C. Vagenas, Phys. Lett. B 503 (2001) 399.

\bibitem{17} Q. Q. Jiang, S. Q. Wu, X. Cai, Phys. Rev. D 73 (2006) 064003.

\bibitem{18} S. Q. Wu, Q. Q. Jiang, JHEP 0603 (2006) 079.

\bibitem{19} M. K. Parikh, Int. J. Mod. Phys. D 13 (2004) 2355.

\bibitem{20} S. W. Zhou, W. B. Liu, Mod Phys Lett A 24 (2009) 2099.

\bibitem{21} X. X. Zeng, Mod. Phys. Lett. A 24 (2009) 625.

\bibitem{22} P. D. Nation, M. P. Blencowe, F. Nori, Landauer transport model
for Hawking radiation from a black hole, arXiv: gr-qc/1009.3974 (2010).

\bibitem{23} S. Iso, H. Umetsu, F. Wilczek, Phys. Rev. D 74 (2006) 044017.

\bibitem{24} R. Banerjee, S. Kulkarni, Phys. Rev. D 79 (2009) 084035.

\bibitem{25} L. Alvarez-Gaume, E. Witten, Nucl. Phys. B 234 (1984) 269.

\bibitem{26} W. A. Bardeen, B. Zumino, Nucl. Phys. B 244(1984) 421.

\bibitem{27} H. Banerjee, R. Banerjee, Phys. Lett. B 174 (1986) 313.

\bibitem{28} R. Bertlmann, Anomalies in quantum field theory, Oxford
Sciences, Oxford, 2000.

\bibitem{29} K. Schwab, E. A. Henriksen, J. M. Worlock, et al, Nature 404
(2000) 974.

\bibitem{30} L. G. C. Rego, G. Kirczenow, Phys. Rev. B 59 (1999) 13080.

\bibitem{31} L. D. Landau, E. M. Lifshitz, Statistical physics (3rd ed),
Pergamon Press, Oxford, 1980.

\bibitem{32} J. B. Pendry, J. Phys. A 16 (1983) 2161.

\bibitem{33} P. C. W. Davies, J. Phys. A: Math. Gen. 11 (1978) 179.

\bibitem{34} W. H. Zurek, Phys. Rev. Lett. 49 (1982) 1683.
\end{thebibliography}
\end{document}